# Beyond Virtual Bazaar: How Social Commerce Promotes Inclusivity for the Traditionally Underserved Community in Chinese Developing Regions

Zhilong Chen[*]
Department of Electronic
Engineering, Tsinghua University
Beijing, China
czl20@mails.tsinghua.edu.cn

Hancheng Cao[*]
Computer Science
Stanford University
California, United States
hanchcao@stanford.edu

Xiaochong Lan
Department of Electronic
Engineering, Tsinghua University
Beijing, China
lanxc18@mails.tsinghua.edu.cn

Zhicong Lu
Computer Science
City University of Hong Kong
Kowloon, Hong Kong
zhiconlu@cityu.edu.hk

Yong Li[†]
Department of Electronic
Engineering, Tsinghua University
Beijing, China
liyong07@tsinghua.edu.cn

## ABSTRACT

The disadvantaged population is often underserved and marginalized in technology engagement: prior works show they are generally more reluctant and experience more barriers in adopting and engaging with mainstream technology. Here, we contribute to the HCI4D and ICTD literature through a novel "counter" case study on Chinese social commerce (e.g., Pinduoduo), which 1) first prospers among the traditionally underserved community from developing regions ahead of the more technologically advantaged communities, and 2) has been heavily engaged by this community. Through 12 in-depth interviews with social commerce users from the traditionally underserved community in Chinese developing regions, we demonstrate how social commerce, acting as a "virtual bazaar", brings online the traditional offline socioeconomic lives the community has lived for ages, fits into the community's social, cultural, and economic context, and thus effectively promotes technology inclusivity. Our work provides novel insights and implications for building inclusive technology for the "next billion" population.

## CCS CONCEPTS

• **Human-centered computing** → Empirical studies in HCI.

## KEYWORDS

social commerce, inclusive, underserved, bazaar, HCI4D

[*]Equal contribution.
[†]Corresponding author.

## 1 INTRODUCTION

The disadvantaged population has long been identified as underserved and marginalized in technology engagement by Human-Computer Interaction for development (HCI4D) and Information and Communication Technologies for Development (ICTD) communities [24–26]. For example, abundant literature has shown that people in developing regions are generally more reluctant and experience more barriers in adopting and engaging with mainstream technology compared with the population in developed regions [23], which can result in non-use of popular platforms (e.g., Facebook [52], Amazon [15]). While there are important infrastructural, economic, and business reasons behind the community being underserved, product design also plays a key role in hindering the adoption and engagement of the underserved population: practitioners often fail to understand the situated needs [4, 49], technical barriers [2, 15], and unique cultural economic context [52, 87] of such communities, thus making platforms "hard to learn" and "hard to use". How to properly design platforms and promote technology inclusion among such communities remains largely an open question.

Here, we contribute to the literature through a novel "counter" case study on Chinese social commerce, where the traditionally underserved community in China – those residing in small towns and villages who are largely underserved and marginalized by earlier IT development (e.g. traditional e-commerce) – are key players [66]. Social commerce (e.g., Pinduoduo[1], Beidian[2]) is a recent phenomenon in China, attracting hundreds of millions of users to make everyday purchases [11, 67]. Social commerce embeds online shopping experience within instant messaging platforms, such as WeChat, thus people could freely share products from online shops with their social circles (e.g., friends and family) - often motivated by economic benefits such as getting a bargain – and make purchases, with the purchased goods shipped to their home[3]. From an HCI4D and ICTD perspective, social commerce is interesting and unique in that it demonstrates a "bottom-up" diffusion process as opposed to the traditional "top-down" process demonstrated by most other technologies: **1) social commerce first prospers in developing regions (e.g., small towns and villages) in China before adopted by the population in large cities**, demonstrating an unusual influence from developing regions and the traditionally underserved community, to developed regions and traditionally well-served community [98], and **2) social commerce has been heavily used and engaged by the traditionally underserved community, with a high retention rate**. In fact, we reveal from our study that many users from the traditionally underserved community enjoy a variety of purchases on social commerce. Statistically speaking, Pinduoduo, the most popular social commerce and the second largest e-commerce platform in China, started by targeting users in third and fourth tier Chinese cities [68] – the widely regarded underdeveloped regions in China, and quickly achieved gross merchandise volume (GMV) of $15 billion only two years from launch, becoming the fastest growing e-commerce in history and demonstrating a good example taking advantage of the so-called "sinking market" [66]. Among 568.8 million monthly active users on Pinduoduo (as of June 2020) [75], over 60% come from cities lower than tier 3, with approximately 36.7% from the fourth and lower tier cities, or traditionally the least developed regions in China [74]. In comparison, on other Chinese E-Commerce giants, active users primarily come from large cities: for example, on JD.com, one other Chinese E-commerce giant, 88% users come from large cities, with only 8% come from mid-size towns, and 2% come from small towns and rural communities as of 2021 [89].

---
[1]https://www.pinduoduo.com/
[2]https://www.beidian.com/
[3]https://en.pinduoduo.com/company

To the best of our knowledge, so far there have been very few cases and studies on successful and inclusive use of "grassroots technology" that targets the traditionally underserved population [50] and eventually achieves global impact. Here we focus on the grounded scenario of social commerce and seek to understand how its design helps engage the traditionally underserved population, *i.e.*, reasons behind its technology inclusivity. Specifically, we ask the following research questions:

**RQ1**: How does the traditionally underserved community[4] in Chinese developing regions engage in social commerce? How do they adopt social commerce, make purchases, and interact with others on the platform?

**RQ2**: How do characteristics of social commerce situate with the traditionally underserved community in Chinese developing regions and promote inclusivity?

To answer them, we conducted 12 interviews with social commerce users from the traditionally underserved community in Chinese developing regions. We carried out in-depth analysis and drew conclusions not only from their direct experience and perception towards social commerce, but also their social, cultural, and economic context, which enables us to position social commerce on a greater scope and understand its central role in the users' daily lives. Based on prior literature on the bazaar economy, we demonstrate social commerce is in many ways acting as a "virtual bazaar" for the population in underdeveloped regions, bringing online the traditional offline social economic lives the community has lived for ages, and thus effectively lowers tech barriers and promotes tech inclusivity. Social commerce is in many ways substituting functions and experiences of "physical" bazaars for them while also adding novel life experience thus beyond a "virtual bazaar". For instance, we show that users adopt social commerce and purchase products through word of mouth among strong ties, meet and hang out with their friends and family online via shopping, and demonstrate reciprocity and clientization in a similar way as "offline bazaars". Moreover, users are able to purchase cost-efficient goods from other places, maintain social relationships (e.g., *guanxi*, *renqing* [78]) through novel social economic activities (even with friends and families living far away) over the platform, which traditional bazaars are hard to offer yet satisfy their customs and benefits. We discuss in detail insights and implications for building inclusive technology for the traditionally underserved population as inspired by social commerce, such as proper use of metaphor and incentive designs that suit the unique cultural and economic context of the community.

## 2 RELATED WORK AND BACKGROUND

In this section, we position our work in the literature of bazaar economy (theoretical basis of the analyses), HCI4D (the line of literature we make primary contributions to), and social commerce (our focal subject). We end this section by illustrating the background of the traditionally underserved community in Chinese developing regions, which we situate our study in.

### 2.1 Bazaars and Bazaar Economy

Originating from the Middle East, bazaars have long been identified as a traditional and essential form of market and economy [34, 36, 37]. Oxford English Dictionary defines a bazaar as "a Middle Eastern marketplace or permanent market, usually consisting of ranges of shops or stalls, where all kinds of merchandise are offered for sale"; Merriam-Webster Dictionary identifies a bazaar as "a market (as in the Middle East) consisting of rows of shops or stalls selling miscellaneous goods"; while Longman Dictionary of Contemporary English uses "a market or area where there are a lot of small shops, especially in India or the Middle East" to define a bazaar.

The academia has been actively characterizing bazaars and the corresponding bazaar economy. Geertz introduced the bazaar economy in Modjokuto, Indonesia and highlighted the fragmentation of commerce flow which consists of numerous unrelated person-to-person transactions as a prominent feature of the economy [36]. He illustrated the essence of three aspects that shape the Modjokuto bazaar economy: 1) good and service flow, where goods were mostly unbulky, with high turnover while small volume per sale and where commodities moved in circles among traders; 2) regulatory mechanisms with a dynamic price system, complex credit relationship balance, and high division of risks and profits; 3) social and cultural specificity. Geertz later delved into the information and search in peasant marketing through analyzing Moroccan bazaars [37]. It was discovered that information was "poor, scarce, maldistributed, inefficiently communicated, and intensely valued" and none of ceremonial distribution, advertising, prescribed partners, or product standardization were present. Social relationships centered around good and service production and consumption,

---
[4]Participants described their information and technology experiences as "traditionally underserved", which we use in the following of the paper.

where stable clientship ties formed between buyers and sellers. Clientelization and multidimensional intensive bargain were recognized as two main procedures in people's information search towards price and quality. Chandra et al. [14] investigated the technology consumption practices in bazaars in Bangalore, India and Bangladesh. Based on practice theory, they identified searching, clientelization, bargaining, and testing as integral market practices and manifested how materiality, relationships, and situated knowledge are functioning through these practices. They highlighted the importance of social capital between customers and buyers in shaping the consumption experiences in bazaars and advocated that new online markets should accommodate customers' comfort and preference ingrained in traditional practices so as to enter the global south better.

Our work builds on these studies to demonstrate the resemblance of social commerce to traditional offline bazaars, and to justify our argument of social commerce functioning as an extended "virtual bazaar". We show how social commerce facilitates experiences beyond bazaars that well-suit the community's social, cultural, and economic context, which further promotes technology inclusivity.

## 2.2 Technology Use in Underserved Community

Our work focus on studying social commerce use in the traditionally underserved community residing in small towns and villages who are largely underserved and marginalized by earlier IT development in China, echoing and contributing to HCI4D and ICTD literature concerning technology use in traditionally underserved communities, such as developing regions [23, 39]. Existing literature has been highlighting the gaps and marginalization of technology usages induced by regions and regional characteristics [23], as well as technological, social, and cultural barriers faced by traditionally underserved communities in adopting technology [44]. For instance, rural areas have been identified as more peripheral compared with urban counterparts [43]; people in the developing regions of the global south are more marginalized and disadvantaged than their peers in developed western countries [8].

Past research has investigated the uniqueness of rural areas and consequent HCI challenges and opportunities. For example, rural areas are characterized by small population size and sparse spatial distributions, where people are geographically segregated, and lack digital literacy and computer skills [44]. They are found to be relatively poor in infrastructural support such as Internet connectivity [63] and lack access to social and health services, which yields design opportunities [53]. That rural economy depends largely on farming has aroused interests in design specifically for farmers and farming [57, 82, 83]. The more localized culture and values that rural people share can pose barriers for technology design to account for [55]. Some other studies aimed at uncovering the differences induced by the urban-rural divide. For example, Gilbert et al. [38] studied the behavioral differences between rural and urban social media users and discovered that rural people have fewer online friends while their friends share higher locality. Other works have indicated rural-urban differences on geographic information systems [45] and peer-produced content platforms [54]. Bearing these discoveries in mind, Hardy et al. underscored the necessity of situatedness in understanding user behaviors in rural areas and recognition of the resilience and resourcefulness of rurality [44].

On a global scale, postcolonial computing literature indicated that power, authority, legitimacy, participation, and intelligibility issues occur when technology diffuses across the globe [49]: what works well in rich developed regions could fail in poor developing regions [49]. To enable better technology inclusion, studies have 1) focused on how traditional practices can inform HCI research and design [14, 85, 86], 2) sought to study the difficulties encountered when developed technologies probe into developing countries and discuss possible solutions [3–5, 15, 52], 3) focused on how specific contexts of the developing countries result in the appropriated usages of technology and design [29–32, 50, 51], and 4) how technology and HCI design changes situated practices of people in developing countries [64, 65, 87].

In sum, prior studies have demonstrated how regional differences can vastly shape users' technology use, which can lead to contingent difficulties as well as possibilities for design, highlighting the necessity of situated investigations. We follow these lines of studies to situate the technology use of the traditionally underserved communities in Chinese developing regions. However, contrary to prior endeavors which mostly focus on how marginalized communities accommodate existing "mainstream" technology with limited users, we seek to explore the "counter" case of social commerce which engages the traditionally underserved community with millions of users and reversely influences "mainstream", effectively acting as one of few examples of successful "grassroots" technology.

## 2.3 Social Commerce

Social commerce is usually referred to as a family of e-commerce that leverages social media/social relationships to boost economic transactions and activities [21, 58, 99]. Laying its theoretical foundation upon the social and economic literature on the embeddedness of economic actions and social structure [40] and the complex interplay between market organization and trading relationships [92], social commerce comes in various forms [21]. For example, social network driven platforms such as Facebook-based F-commerce and Twitter-based T-commerce gives birth to social commerce through incorporating commercial features [20, 58]. As such, existing social networks on these platforms can be made good use of for facilitating transactions [20, 72, 97]. Traditional e-commerce platforms such as Amazon, also develop social commerce through incorporating social media features [58]. Novel social commerce platforms with distinct mechanisms emerge as well, e.g., Groupon [47].

Abundant research attempts have been made to better understand social commerce [10]. On the commerce side, past research has widely examined how factors such as social support and uncertainty [6], social presence [59], social desire [56], social climate [88], trust of the institution [60], social commerce constructs [42], etc. influence user's purchase intentions. Within the realm of HCI and social computing, a line of past research has been dedicated to uncovering the unique user experiences on social commerce. For example, Cao et al. [12] examined the network characteristics of a typical social commerce site, demonstrated its topological decentralization, rapid growth via invitation cascade, and high conversion rate related to user proximity and loyalty. Cao et al. [11] revealed how intimate social relationships shape the unique user experiences and purchase decision-making processes of instant messaging based social commerce. Xu et al. [93] proposed the mechanisms of better matching, social enrichment, social proof, and price sensitivity on social commerce, and validated their effectiveness in helping people make purchase decisions, and Piao et al. [73] identified how social commerce alleviates algorithmic homogeneity by bringing friends into the recommendation loop. Xu et al. [94] further analyzed the mechanisms behind acceptance of intermediary/agent invitations in intermediary-mediated social commerce, while Chen et al. [17] probed into how these intermediaries could make the best of their roles.

We build upon these studies to pursue a more profound understanding of social commerce. We follow the line of studies in HCI and social computing, and situate our study into WeChat-based social commerce (e.g., Pinduoduo, Beidian) in China, where users share product links from social commerce sites over WeChat groups with friends and family members, motivated by either monetary incentives (e.g., discounts) or internal motivations (e.g., recommending products with pleasant experiences) [11, 12]. Through group chats and direct messages (DM) over WeChat, useful product information is shared among users, thus social activities and economic behaviors are highly embedded [11, 12]. However, different from prior endeavors which focus on overall experiences on these platforms, we pay attention to how social commerce provides experience fitting the social, cultural, and economic context of traditionally underserved people, and promotes technology inclusivity.

## 2.4 Chinese Developing Regions and Traditionally Underserved Community

The focal population of the current study reside in Chinese developing regions with city tier below tier 3 [68], which accounts for a large proportion (50.3%) of the population in China [16]. Despite the size of the population, they are traditionally underserved and marginalized from mainstream technology, e.g., only 10% users of the Chinese leading e-commerce JD.com come from regions other than large cities. Part of the reason for being traditionally underserved is the community's relatively low purchasing power and poor infrastructure: the average GDP per capita of tier 3 cities is 31,000 CHY compared with 54,000 CHY GDP per capita of tier 1 cities in 2014 [22]; GDP per capita for smaller cities and rural regions is even lower. The community also accounts for lower literacy. As of 2019, around 73% of Chinese who lived in lower-tier cities and rural areas did not receive any tertiary education, and only less than 8% of the lower-tier consumers have acquired a bachelor's degree or higher [1]. Thus, they may naturally find it harder and less confident to learn new technology [90, 100].

Fortunately, the community has experienced rapid development in recent years in economy and infrastructure. Smartphones [69, 70], Internet access [13, 70], and delivery services [33, 71] have penetrated into these developing regions, and mobile payments such as WeChat Pay have become largely available [79, 81]. Thanks to the development, recent years have seen this community adopt technology such as instant messaging apps (e.g., WeChat) following the population in more developed regions, although their uses are mostly limited to basic functions such as communications. It is not until very recently when a new generation of applications, such as social commerce, first becomes common practices within the community, then penetrate and influence the more developed regions and the traditionally better-served population in a direction reverse to the traditional technology diffusion process [66].

Moreover, it is worth mentioning that the community is strongly influenced by traditional Chinese economic and cultural practices [62]. Economically, the community is used to shopping in markets or bazaars [46] as modern supermarkets have only entered their horizons in recent years. They are usually not financially well-off and have relatively limited budgets [101], which leads to their highlights on cost-efficiency of purchases rather than brands, and sensitivity to small price reductions [46]. They would quite often bargain with sellers, and compare different products so that they could make the most cost-efficient decisions (e.g., through exhaustive searching and haggling), where there is generally not a transparent price for products and people would use bargaining to reach a consensus. In comparison, in larger cities where the population is much better served by IT technology, people are used to shopping branded products either in offline supermarkets or online e-commerce such as JD.com or Alibaba, and bazaars are hardly seen in recent years. Socio-culturally, the focal community of the current study has been regarded as more dominated and reflective of the traditional Chinese culture [62], including collectivism [95, 96], *mianzi* [77], *guanxi* [9, 78], and *renqing* [78], whereas communities residing in larger cities are relatively much more influenced by western, individualism, and capitalism culture. We will elaborate in findings how social commerce's adaptation to the unique characteristics of our focal community contribute to their willing engagement.

## 3 METHOD

**Participant recruitment.** To delve into our research questions, we conducted a qualitative study through in-depth interviews with social commerce users from the traditionally underserved community in Chinese developing regions. To achieve the research goal, we recruited participants who satisfy the following criteria: 1) they have been using at least one social commerce platform for more than three months, 2) they live in county towns or rural areas of Chinese cities with no higher than fourth tier [68], 3) their highest level of education is no higher than senior high school or technical secondary school, and 4) their self-reported IT self-efficacy is no higher than 3 on a 5-point scale. We adapt from prior literature [18, 48] to measure their IT self-efficacy by probing their confidence in understanding how IT works and learning IT applications by themselves. We utilized these criteria because 1) city tier plays an important role in the Chinese context, where metropolitan areas are traditionally more central in technology consumption while low-tier cities are more peripheral and less developed [90, 100], 2) people with poorer education background are traditionally more disadvantaged in technology adoption, and 3) people with lower IT self-efficacy generally find it harder to adopt and use technology [48]. We leveraged our social relationships to approach people in these communities and used snowball sampling [7] to 1) ensure that participants satisfy the aforementioned selection criteria and 2) guarantee interview quality assuming that inherent trust among recommended participants [80] would lead participants to better reflect their authentic experiences. Our work has been reviewed and approved by the local Institutional Review Board.

**Interview procedure.** We conducted our interviews through remote audio calls between May and August 2021 in a semi-structured manner. We investigated how participants adopt social commerce, what enables them to make purchases on social commerce, what they purchase on social commerce, how they make purchases on social commerce, and why they prefer making purchases on social commerce, etc. We constantly asked them to compare social commerce with offline shopping and the use/non-use of traditional e-commerce. We also focused on how characteristics of social commerce suit the social, cultural, and economic context, and satisfy the needs of our participants. We followed the grounded approach [19] to interview and analyze our contents iteratively and consistently revise our interview protocol for better theme extraction. We continued interview data collection until we reached theoretical saturation [19], where the incorporation of new data would not induce new themes. In this way, we executed 12 interviews. Each of the interviews lasted for 30-60 minutes and was compensated for 50 CHY. The interviews were all conducted in Mandarin and were audio-taped after we received oral consent from the interviewees. We further transcribed the recordings of the interviews with the combination of automatic transcription service and manual rectification. To better protect participants' privacy, we removed all identifiable and sensitive information.

**Interviewee details.** Table 1 shows the detailed information of our interviewees. According to our participants, the vast majority (75%) of them have seldom or never used computers and just started to use IT applications in recent 1-2 years with smartphones. Their self-reported IT self-efficacy is relatively low because of their low confidence in comprehending and mastering IT applications on their own. The community is referred to as "traditionally underserved" because our participants chose to use the term to reflect their experiences for information and technology usages, where they used "traditionally" to denote before the adoption of technologies such as social commerce platforms. As expressed by them, the psychological sense of distance from these applications is a major factor contributing to this traditionally underserved status, which has been largely overcome by recent social commerce as we will show.

**Data analysis.** We adopted open coding [19] for data analysis. Two native Mandarin-speaking authors coded the first 20% of the interview transcriptions separately and came together to discuss the codes until there were no disagreements. One of them then coded the rest of the transcriptions and routinely discussed with two other Mandarin-speaking authors to address uncertainty on the codes. This author then translated the codes and corresponding quotes into English and the translations were further validated by those two other Mandarin-speaking authors. The research team then gathered together to discuss the contents derived through prior processing. We developed and repeatedly revised the themes emerging from our data through sub-categorization and constant comparison [84].

## 4 FINDINGS: HOW THE TRADITIONALLY UNDERSERVED COMMUNITY ENGAGES ON SOCIAL COMMERCE

Most of our interviewees from the traditionally underserved community in Chinese developing regions have been deeply engaged by social commerce and could now *"spend several hours a day on these platforms"* (P3, P9, P11). Some users do not use other e-commerce at all even if they have heard of them, or quit other platforms after encountering barriers at early stages (P3). Purchases on social commerce platforms can take up *"the vast majority of daily consumes"* (P2, P9, P11), all demonstrating the heavy use of social commerce in our focal community.

In fact, social commerce take the place of traditional offline bazaars to a great extent within the community. As reflected by our interviewees, some sharply reduce their purchases offline and move them online to social commerce platforms: social commerce platforms become their *"first choice when lacking certain items"* (P2) or even *"almost only buy from here (social commerce platforms)"* (P3) rather than going to the traditional offline bazaars and markets. As reflected by P3,

> "I can get what I want from [social commerce platform name]. Therefore, I am not willing to go to offline shops and bazaars for shopping anymore ... I just buy them from [social commerce platform name] with my smartphone." (P3)

To understand what cultivates participants' strong preferences towards social commerce platforms from a design perspective, we investigate: 1) how participants adopt social commerce (Section 4.1), 2) what products participants consume on social commerce (Section 4.2), 3) what purchase behaviors (Section 4.3) and social experiences (Section 4.4) exist on the platforms. We lay special focus on how these elements fit the characteristics of the traditionally underserved community in Chinese developing regions. A general theme that emerges, which we will discuss in detail in Section 5, is social commerce is in many ways acting as a "virtual bazaar" where the community's familiar experiences of offline bazaars are moved online – reflected in the adoption process, products being transacted, and user social economic practice. Beyond resemblance to the bazaar economy, social commerce provides convenience and novel social buying experience that fits the community's benefits and context, which further engage users and promotes technology inclusion. While certain elements in Section 4.1 and 4.2 may be shared by some e-commerce platforms/other communities, unique experience provided by social commerce discussed in Section 4.3 and 4.4, and the combined power of all factors mentioned here make social commerce especially appealing to our focal community.

### 4.1 Adoption Backed by Strong Ties

Social commerce platforms leverage the power of strong ties to adopt users, and have proved to be especially effective with the traditionally underserved community, which can relate to not only *how* users are introduced, but also *why* they decide to use the platforms according to domestication theory [15].

**Getting to know the platforms: introduction from strong ties.** Prior study identified online and offline advertisements as major sources that help people get to know e-commerce platforms in developing regions [15]. However, this is not the case for the adoption of social commerce. Similar to bazaars where recommendations from peers are decisive in getting new customers on board [14, 37], interviewees most frequently get to know social commerce platforms through recommendations from their strong ties such as friends (e.g., P2, P4, P8), or relatives (e.g., P3, P5). For example, as narrated by P5,

> "My sister used [social commerce platform name] earlier. It was that my sister and my sisters-in-law used it first and recommended to me that I began to know about it and use it." (P5)

Table 1: Basic Information of Interviewees.

| Id | Gender | Age | City Tier | Education | IT Self-Efficacy | Id | Gender | Age | City Tier | Education | IT Self-Efficacy |
|---|---|---|---|---|---|---|---|---|---|---|---|
| P1 | F | 40-50 | 4 | Senior High School | 2 | P2 | F | 70-80 | 4 | Junior High School | 1 |
| P3 | F | 70-80 | 4 | Junior High School | 1 | P4 | F | 30-40 | 4 | Technical High School | 2 |
| P5 | F | 50-60 | 4 | Junior High School | 1 | P6 | F | 60-70 | 4 | Technical High School | 2 |
| P7 | F | 50-60 | 4 | Junior High School | 1 | P8 | M | 50-60 | 4 | Technical High School | 3 |
| P9 | M | 60-70 | 4 | Senior High School | 2 | P10 | M | 40-50 | 4 | Technical High School | 2 |
| P11 | F | 30-40 | 4 | Junior High School | 2 | P12 | F | 30-40 | 4 | Technical High School | 3 |

This high reliance on strong ties accords with the traditional Chinese collectivism culture, where kinship and strong ties play a foundational role in shaping people's behaviors [35]. Compared to advertisements, their words are regarded as more trustworthy and thus are more likely to be taken into consideration:

> "It depends on recommendations. When it is widely recommended among relatives and friends, it becomes popular ... If it were not the recommendations by strong ties, I would not use it. Recommendations by people not so familiar are not that trustworthy." (P3)

Backed up by strong ties, users tend to pay more attention to these trustworthy recommendations and follow them, which is especially important for the traditionally underserved communities. Traditionally, even though some technology (e.g., e-commerce) are pervasively advertised, community members regard them as distant or irrelevant and pay little attention to the advertisements due to their relatively limited digital competence and the fact that no people they know use the technology. In comparison, introduction and recommendations from strong ties successfully intrude their horizons through social relationships. As such, these communities can more easily get to know about the platforms.

**Getting to use the platforms: peer usages and help from strong ties enhance confidence of competence.** Prior work on typical e-commerce adoption indicated that digital illiteracy can lead people to be conservative or even reluctant to adopt online shopping and result in non-use [15]. Such conservation and reluctance is also manifested by some of our participants when they describe why they do not intend to try some other e-commerce platforms. For example, here is an excerpt of our interview with P2:

> P2: "Other people tell me about [other e-commerce platform name]. But I don't know how to use them."
> Q: "The mechanisms of [other e-commerce platform name] is just similar to [social commerce platform name]."
> P2: "I am old. My brain is not functioning so well. I think I may not be able to handle them. [Social commerce platform name] is enough."

Similar circumstances are also reflected by other interviewees such as P3 and P6. For them, their relatively low digital literacy results in a lack of confidence in their digital competence. For fear that their digital competence would not qualify them for using new technology, they hesitate or even refuse to take their step further.

However, for social commerce, the adoption triggered by the strong ties' reference alleviates the obstacles to a large extent. Introduction from people they know, especially those of similar age and with similar backgrounds (thus similar level of digital literacy), can be extremely effective in persuading these individuals to adopt the platform. Usage by their friends creates a sense that these platforms are within their competence and that they can handle the platforms. As such, their reluctance and resistance towards the adoption of the platforms are resolved:

> "My elder sister first introduced [platform name] to me. She could use it, then why can't I? (laugh) Introduction from people of my same age made me feel confident of my competence ... It made me feel that it is something that I can master and created a sense that I am qualified for usages." (P5)

However, some interviewees could still encounter some obstacles at the initial start of their usages. For them, better-educated family members and friends of similar age who have already begun to use the platforms, can be an especially important source of help:

> "At the very beginning, you need to have these relatives and friends to tell you and accompany you. They tell you and show you how to operate the app. Others tell me, and then I tell someone else. When one tells another, the snowball is rolling bigger and bigger. At first, I thought it might be too difficult or something. But in fact, if acquaintances teach me a bit, I begin to think that it is not that difficult indeed." (P3)

Such help can be especially important for our participants from the traditionally underserved communities: rather than official guidelines which they find hard to follow, help from peers better situates their difficulties and lends them confidence of their competence to master the operations on the platforms. As a result, they gradually get familiar with and accustomed to the usage of the platforms. Such adoption mechanism thus distinguishes itself from traditional word-of-mouth where people share and adopt out of interest/monetary incentives.

In sum, similar to the offline adoption process in the bazaar economy (*i.e.*, the community would go shopping at a certain place if they get referred by their close relationships), the traditionally underserved community in Chinese developing regions adopt social commerce with strong ties' influences and help in a collective fashion, effectively lowering the "entry" bar they may experience on other technology.

## 4.2 Virtual Bazaar for Daily Purchases with Diverse and Cost-efficient Choices

Social commerce has enabled our participants to purchase items not only similar to what they can get in offline bazaars, but also more suitable to their preferences and benefits, which contributes significantly to their willingness to engage on these platforms.

**Cost-efficient daily necessities from a variety of online shops.** Participants buy a wide range of products on social commerce similar to bazaars. For example, as noted by P2,

> "There are so many things that I buy from [social commerce platform name]! From clothes, shoes, food, to daily necessities such as towel, toilet paper, soap, shampoo ... All of them can be bought from it. [social commerce platform name] was amazing." (P2)

Similar perspectives are also frequently mentioned by other participants (e.g., P3, P5-P9). For them, *"small items such as pots, pans, bows, socks, slippers"* (P1), *"miscellaneous goods"* (P8) or *"items that cost only several CHY or several dozens of CHY"* (P9) are mentioned as the main constituents of their purchases on social commerce, which matches exactly the items most often sold over bazaars. With access to a variety of products, people can almost *"buy everything (from the platform) at the same time"* (P9), which brings much convenience.

Social commerce is also praised for its cost-efficiency. Compared with traditional e-commerce platforms, one essential aspect that enables social commerce to further lower down prices is its adoption of the group buying format: buyers can form a group to buy the same product so as to enjoy a lower price (P10). Through *"lowering prices but increasing turnovers"* (P8), these platforms enable more plausible prices that fascinate these audiences while guaranteeing the overall profits. Such cost-efficiency is particularly appreciated by the traditionally underserved community in Chinese developing regions as most people residing in these regions do not earn much financially. Even for those who are better off, most have experienced hard times in the past. As a result, they develop the habit of budgeting strictly and can be sensitive to small price reductions. Products in social commerce suit these people's appetite in providing lower (e.g., P1 & P12) and *"more reasonable"* (P5 & P9) prices. This can be especially important when their daily accesses to offline shopping are limited to a few options, where the prices of some items can be higher (P5).

Moreover, products from small virtual "shops" take up the vast majority of purchases on social commerce platforms, which aligns with bazaars that consist of small shops. Contrary to traditional e-commerce where customers buy mostly branded products [15], a vast majority of items that interviewees buy from social commerce platforms are unbranded items from small, non-famous shops integrated on the platforms (P1, P6). Their relatively limited budgets prevent them from chasing brands that are often with higher prices. Instead, they opt for the most cost-efficient ones with high utility, which they believe they can make the right judgments based on their life experiences:

> "For people like me, we mostly buy daily necessities (on these platforms). We make a living ... Most of these items are not with famous brands and from small and less-known shops. We focus on the utility. More or less we have a degree of judgments based on our life experiences ... At least I haven't encountered any unpleasant purchase experiences in this way." (P3)

**"Products you would not get or even know from local".** Social commerce further enables the traditionally underserved population in developing regions to approach items not available locally. Traditionally, most people in the community make purchases (especially daily necessities which they now buy through social commerce) from local shops, bazaars, and markets, the item coverage of which is relatively limited compared with major cities: *"in big cities you can buy all kinds of things, but here (in developing regions) the range of choice is limited"* (P12). However, with aforementioned easy adoption and infrastructural support, social commerce provides them with the competence of online shopping and makes it possible for them to equitably reach a wider variety of items. As a result, they manage to *"have a wider range of choices"* (P9), helping them access more options of cost-efficient products.

Social commerce also enables the community to approach competent items previously unknown. For example, as illustrated by P9,

> "You don't even know about some items before. Last time I bought the welding glue. It only cost one CHY, but can stick everything. It was very convenient and could stick things solidly. It is a lot more useful than the 'small white glue' that I used to rely on." (P9)

In short, social commerce is functioning the role of bazaars, providing bazaar-like products to the traditionally underserved community in developing regions with more options in a cost-efficient manner, which greatly appeals to the community.

## 4.3 Online Purchasing Behavior Developed from Offline Bazaar Practices

As expressed by interviewees, social commerce also demonstrate interesting economic behavior. Consistent with Chandra et al.'s [14] highlights on purchase practices of searching, clientelization, bargain, and testing in offline bazaars, we discover that similar practices emerge on social commerce. As a result, social commerce platforms accommodate the community's offline economic customs while bringing new merits.

**Searching.** Searching refers to the attempt to find products of interest, which is the preliminary step leading to actual purchase. In offline bazaars, searching is completed in the physical space, where physical distance, as the cost of searching, would influence how people make decisions [14]. Purchases on social commerce vastly reduce search costs through substituting the journeys that require *"taking a car to 20-30 kilometers away"* (P6) by browsing, sweeping, and clicking on fingertips (P3).

However, challenges for searching are encountered along the way as well. A larger number of products become available on social commerce compared to offline settings, which could lift people's burden in searching (P2, P6). This could be especially prominent for the traditionally underserved community who are not so familiar with the platform. Fortunately, in line with circumstances of offline bazaars where experiences from social networks are used for lessening searching costs [14], the community seek for friends' sharing to address the information overload in social commerce. As explained by P2,

> "There are so many items on [social commerce platform name] and you can get overwhelmed. Sometimes you have no idea which one to choose. However, when my friends buy something good and cost-efficient, they share it with me and recommend it to me. It would save a lot of my energy to follow them." (P2)

As shown by P2, recommendations from peers can significantly ease the process of searching. Sometimes users may not be familiar with certain items or the sales of certain items, but their friends could be. From friends' recommendations, others' knowledge and past experiences can be well utilized as trustworthy sources of references[5]. With their recommendations, users will no longer suffer from choice overload and thus the purchase experiences are greatly improved. Such mechanism functions notably well when users' detailed knowledge of items is relatively limited, which is often the case for the traditionally underserved community in Chinese developing regions.

**Clientelization.** Clientelization in the bazaar economy underlines the consumer-seller relationships through interaction repetition and social capital accumulation [37]. Parallel to prior work on offline bazaars which demonstrates the indispensability of reciprocal and symmetrical customer-seller relationships on retaining repetition [14], we find a similar mechanism on social commerce, where pleasant experiences strengthen people's willingness to repurchase from the same seller. For example, P1 would *"add a seller to favorite when buying something satisfying from the seller and choose from him/her first next time when needing something new"* (P1).

For others, their sources of purchases are more diverse and they possibly *"do not have specific sellers to stick to"* (P9), making the customer-seller connections shallower than those of traditional bazaars. Nevertheless, they enunciate similar forms of clientelization on the platform level. For example, as P3 specifies,

> "I am accustomed to using [social commerce platform name] ... I only trust [platform name]. My relatives, friends, and I are all using it and haven't encountered any mistakes in our years of purchases ... I have never thought of trying other similar platforms. [social commerce platform name] is enough for me." (P3)

On social commerce platforms, our participants' prior satisfactory experiences as well as their strong ties' cultivate users' trust in the platform. When these purchases are pleasant over and over again, their preferences over the platform accumulate although the online shops they visit may differ. A positive feedback loop is thus formed, where repetitive interactions between these users and platforms constantly strengthen the customer loyalty to the platforms, effectively sustaining a high retention rate. This mechanism appears especially powerful for the traditionally underserved community whose intention to try out other platforms is hampered by their relatively limited confidence of technology competence.

**Bargain.** Bargaining, which means price negotiation between buyers and sellers so as to reach an acceptable consensus, is regarded as an essential characteristic of purchase practices in offline bazaars [14]. Bargaining on social commerce is ubiquitous as well, although may deviate from the classical offline-bazaar bargaining procedure in forms. As noted by P8, they have been accustomed to bargaining:

> "You have to bargain (on social commerce platforms) ... I like bargaining and have developed the habit of bargaining no matter online or offline." (P8).

In response to their habits, some social commerce platforms have introduced a new kind of "bargain" feature:

> "You send your 'bargain' link to your others, for example, your friends. They click your link to advance your process of 'bargain' ... When your 'bargain' progress achieves 100%, you can finish your 'bargain' and can buy your product at an extremely low price or even for free." (P6)

Similar uses of "bargain" are frequently mentioned and experienced by other participants (e.g., P2-P5, P9, P12). Although these actions under the metaphor of "bargain" sometimes vary from people's traditional haggling practices, our participants show high adaptability to this new form. Firstly, both new and traditional forms of "bargain" aim at purchasing at the lowest price. Considering the community's relatively limited budgets and high sensitivity on price reduction, they have the innate motivation to conduct "bargain". Secondly, both new and traditional forms of "bargain" call for social efforts to achieve the deal: in the traditional bargain, it is a social activity between sellers and buyers [14], while in the new form of "bargain", it is a social activity with peers, especially close relationships such as kinship and real-world close friends. For example, P3-P6 all only reach out to their best friends and families for help on "bargain". As such, although bearing a certain degree of differences from the traditional practices of bargaining, "bargains" on social commerce are quickly understood and fluently practiced by interviewees.

Bargaining also comes in other self-developed forms. Although the publicly marked price system seems hard for customers to conduct traditional bargaining, some users have accommodated their own practices to achieve their goal while meeting the requirements of the platforms. For example, here is an excerpt from P8 who describes how they negotiate prices with the customer service staff of the shops:

> P8: "Sometimes I bargain with the customer service staff, asking if the item could be cheaper. If the staff refuses, I would tell them that in other shops the items could be cheaper than them. If another shop is really cheaper, I would send him/her a screenshot. Sometimes they would send you an internal link with lower prices. Sometimes they would promise you to give you a red pocket back with a certain amount of money afterward."
> Q: "Would the staff really give back the money they promise?"

---
[5]In comparison, traditional e-commerce leverages recommendation algorithm to alleviate the burden yet may be hard for people to trust, especially in our focal community.

> P8: "Yes, he/she would. If he/she doesn't keep the promise, we can just choose to return the goods and the after-sale service can guarantee that we get all our money back."

As described by P8, some users have developed accommodated practices of bargaining in online scenarios of social commerce. The availability of some new technologies, for example, screenshots as mentioned by P8 in the excerpt, makes the procedure easier. The utilization of certain functions, e.g., internal links and paybacks through virtual red pockets over WeChat, helps bargaining work within social commerce. As such, they have successfully moved online their offline negotiation strategies.

**Testing.** In offline bazaars, people evaluate goods or services through testing, *i.e.*, trying them to find out how well they work [14]. People's quest for quality products gives rise to the important role of testing in offline bazaars [14]. However, prior studies on the domestication of e-commerce in the global south indicated that the physical segregation of sellers and customers makes traditional touching and feeling of items seemingly difficult and can hinder usages of the underserved people [15]. In social commerce, the issue is addressed through references of friends' experiences and after-sale services. Specifically, although users themselves cannot test the items in person when making purchase decisions, they can depend on their friends who have made identical purchases and refer to their testing experiences. For example, P5 describes,

> "Sometimes others tell me that they have bought something quite well ... These recommenders are mostly trusted friends and relatives that I get along well with. I have no doubt about them. I just buy whatever I need in the recommendation list without hesitation." (P5)

These circumstances are frequently mentioned and praised by our participants. Although similar reflections could also be found in the comment section, they sometimes *"do not believe in some of the comments"* (P5) because of possible misbehaviors such as click-farming. However, when the recommenders are strong ties, participants' trust in the socially close recommenders leads them to believe in the authenticity and trustworthiness of the evaluations. They just count on these reflections on the items and no longer need further testing on their own.

When these accountable prior purchase experiences are unavailable, our participants also develop modified versions of testing empowered by the impressive after-sale services. For example, as P9 specifies,

> "The after-sale service of [social commerce platform name] is very considerate and convenient. When an item arrives, you check it. If there is something wrong or it does not suit you, you can freely return the item or ask for a replacement in seven days ... This allows us to have a sense of what the item is really like before actually accepting the items. Our trials can be thus made more freely." (P9)

With convenient returns and replacements enabled by after-sale services as backings, users can freely choose items and test upon the items' arrival. Compared to offline bazaars, the aforementioned scenario is only different in that it separates testing from the decision-making process on initial purchases, and delays testing to the item reception phase. As such, the community's familiar way of testing practices is retained to a large extent and their needs for quality purchase are satisfactorily met. Furthermore, interviewees speak highly of the new testing practice as they could change their mind after they make the initial purchase: it reduces impulse buying and saves money (P10), which can be especially detrimental considering the community's sensitivity with price and longing for cost-efficient products.

As can be seen, social commerce acts as a virtual bazaar where users easily adapt their offline bazaar shopping practices that highlight social interaction, in comparison to the recommendation algorithm-driven experience powered by other e-commerce platforms where users hardly interact with peers (especially those they know). Such design of social commerce situates into the community's social cultural background thus appeals to the community. In the following section, we will show how such design gives rise to novel experience that further satisfy the community.

### 4.4 Social Interactions in Virtual Bazaar Shaped by Chinese Socio-Cultural Norms

Social commerce demonstrates interesting social interactions that suit the community's cultural context. Past research shows that economic transactions in offline bazaars deeply embed with social relationships [14, 37], where social interactions between customers and sellers play a crucial role. While seller-buyer interactions are not as salient on social commerce, interviewees demonstrate novel social interactions shaped by several traditional Chinese social cultural norms, such as collectivism, *mianzi*, *guanxi*, and *renqing*, which weigh much more influence over our focal community than population residing in larger cities. Their behavior over social commerce effectively extends the embeddedness of economic and social interactions in new formats beyond the traditional offline setting of bazaars.

**Social interactions shaped by collectivism.** With a collectivism culture, Chinese people have the tendency of concentrating on interdependency and sociability and referring to other peoples (especially strong ties) for decision making [28, 95, 96]. Our participants, who are generally quite influenced by the traditional Chinese culture, highlight how they could socialize with their friends over social commerce, which moves offline "social" shopping online.

For instance, in traditional offline settings, people in Chinese developing regions go shopping with close friends, neighbors, and relatives as a way of maintaining their relationships:

> "Sometimes I go into the street for shopping with friends and neighbors I get along well with. When I see something I really like and when they say that the clothes suit me, I would like to buy one." (P2)

Social commerce enables similar experiences online, whether users are physically co-located or distant. When users are co-located, as mentioned by P5,

> "Sometimes when I am with my friends and relatives, we would go through [social commerce platform name] together. When we see something good, we would analyze it together and would sometimes purchase it together in groups." (P5)

As such, people share their shopping experience together, which helps them make more reasonable decisions. Although similar cases could happen in traditional e-commerce, the group-buying mechanism further incentivizes collective purchasing: it encourages the formation of groups so as to enjoy a lower group-buying price together.

When users are not physically co-located, these actions of social interactions shaped by collectivism can still be present on social commerce. As mentioned by P6,

> "Sometimes when we are not together with each other, we send each other links, saying 'I am thinking highly of something and would like you to see it'. Sometimes it is sent for their advice, while sometimes for letting them know (about the item)." (P6)

As such, others can leave their comments and concerns in chat groups (as social commerce is embedded in instant messaging applications), which maintains the sense of closeness through a distant, asynchronous way that breaks the spatial and temporal restrictions. Such phenomenon can help maintain interactions among physically scattered strong ties and the sense of togetherness. Therefore, social commerce provides a valuable channel to bring our participants together with strong ties who are far away. This is especially important when some of their friends and kinship go to metropolitan areas to seek for higher-paid job opportunities (P6, P12).

**Social interactions shaped by mianzi.** *Mianzi*, or face within the Chinese context [9, 61], represents prestige or reputation. Chinese people have the tendency of trying their best to acquire and maintain *mianzi* and could behave for *mianzi* even if something does not give them direct benefits. Our participants reflect that social commerce enables novel social interactions that help maintain *mianzi*. For example, P5 shares:

> "I stay at home all day long and do not have much to do. Then I just browse these items on [social commerce platform name] ... When I manage to find something good with my efforts, I would like to share it with my friends and relatives, say, 'you see, I have bought something good', to show off. If its price is quite good and its quality is excellent, it would be hard to describe how happy I would be. I would like to tell all my relatives and friends and feel a strong sense of mianzi." (P5)

Similar circumstances are also mentioned by several other participants, where they send out product links in exchange for others' recognition of their efforts in finding cost-efficient products. This can be especially fulfilling when they have plenty of time and have relatively few meaningful things to do: this act of sharing signifies that their spare time is not boringly wasted. They have made good use of their plentiful time and earn recognition. Moreover, their ability to discover those "good" items shows their competence in explorations and identifications. As such, through sharing the information with others, especially with strong ties, they feel that they can acquire *mianzi*.

*Mianzi* can be further enhanced when users are able to get something good at an unusually low price. Through manifesting their capability of achieving the expected utility with the least cost, a sense of superiority can arise. As a result, they would like to show off and are eager to share the information with strong ties, which consequently cultivates a strong sense of *mianzi*. As such, different from prior studies where people mainly show prestige and *mianzi* through showing luxury and wealth [61], social commerce demonstrates the possibility of earning *mianzi* in a "penny-wise" way, which motivates the community to use the platform. Such motivation can be quite powerful given the community is not wealthy financially and hard to afford luxuries to show off for earning *mianzi*.

**Social interactions shaped by guanxi.** *Guanxi* [9, 78], which is typically referred to as connection or "interdependent relationship", is highly valued in the Chinese society. For Chinese people, *guanxi* can function as a variant form of social capital, where good *guanxi* denotes moral obligations and emotional attachments that motivates people to offer others favor [78]. Our participants have also underscored the role of *guanxi* in their social interactions on social commerce platforms. The platforms incorporate features where users could use their social capital backed by *guanxi* in exchange for economic benefits. For example, to achieve a lower price, one needs to ask other users to click their links to be able to purchase the items at a reduced price (P7, P12). To gain virtual and actual rewards in the built-in games supported by social commerce apps, one needs to invite others to click their links for help (P2-P6). Some platforms even provide monetary revenues if deals are successfully made through the special links one shares (P10). As such, social commerce provides possibilities to materialize and make good use of *guanxi*. To receive those economic benefits, users need to find the right audience to ask for help. In Chinese developing regions where *guanxi* plays an essential role in social lives, people rely heavily on *guanxi* to determine whom they turn to, which in most cases are their strongest ties. For example, as articulated by P3,

> "We only choose my close kinship and best friends. It is too embarrassing to bother other more distant ones for help. They click it for you. You get benefits. But they don't. We can't do this to the unfamiliar ... If someone I get along well with asks for my help (on clicks), I would happily help them. They are my relatives and good friends! Why don't I help them?" (P3)

Similar perspectives are also frequently mentioned by other interviewees. They have recognized that not only would such favor-asking requests be disturbing, but its one-way rewarding nature can be embarrassing: only the requester receives benefits themselves, while the helper gains no direct benefit. As such, they regard it inappropriate to make these interactions with people they are not familiar with. However, when the social relationship or *guanxi* among people is strong enough, the social capital underlying strong *guanxi* can compensate for the disturbance and the embarrassment. As illustrated by P3, for relatives and friends, a huge amount of social capital has already been accumulated within their *guanxi* – enough for most exchange of favors,let alone the simple action of clicking. From the helper side, they could also be glad to offer help to their strong ties: it can demonstrate their utility to others and act as a way to gain social capital from their familiar requester. Moreover, such actions of asking for and offering help on social commerce platforms could even help them enhance *guanxi* in turn. As mentioned by P12, these actions can *"facilitate the communications of the ones whose relationships are close but lacks a topic for chatting"* (P12). With the belief that *"guanxi needs to be managed for long-term maintenance"* (P2), these actions are regarded as helpful for keeping the strength of the underlying ties.

**Social interactions shaped by renqing.** *Renqing*, *i.e.*, exchange of favor or reciprocity, is also identified as essential in determining our participants' social interactions on social commerce platforms [78]. For Chinese people, *renqing* is crucial for good *guanxi* [78]: the traditional Chinese culture underlines the importance of interpersonal reciprocity of giving and returning for the maintenance of *guanxi*/relationships. Such pursuit of reciprocity is also mirrored in real-world practices of the social interactions on social commerce as reported by our interviewees:

> *"If they have helped us 'bargain', we would definitely help them too (when they send their links). This is reciprocal. It doesn't make any sense if we do not help them then."* (P4)
>
> *"You send me your link, and I send you mine. You help me a little, and I help you back. We Chinese people stress that these actions are mutual: you owe me a favor, and then I owe you a favor; when things like this happen time after time and we cannot tell how much we owe each other, we become closer friends."* (P6)

As expressed by participants such as P4 and P6, since favor-asking requests mostly bring direct economic benefits only to sharers, they feel that they owe others a *renqing*. As such, they are eager to pay it back when possible because of their innate emphasis on mutuality. Such highlight on reciprocity and mutuality of *renqing* is manifested in interviewees' reluctance in reaching out to people whom they find it hard to help back. For example, P2 says,

> *"You can ask the familiar ones occasionally for help and they would help you because of your guanxi (with them). But it is inappropriate for you to ask them frequently when it is hard for you to equally pay them back. You will owe them favors. So the ones I frequently interact with and ask for help are my relatives and friends who also use [social commerce platform name]. We can easily repay each other."* (P2)

Such phenomenon fits the cultural context of the community where the exchanges of *renqing* are considered crucial: if they cannot pay the *renqing* or favor back through the same actions, they deem it vital to find other ways to compensate the *renqing* or favors. However, it is hard to measure how large such a *renqing* is and *renqing* of what kinds would be suitable to make up for the one they have owed. To avoid complexity like this, social commerce users simply limit their frequent contacts to those they can easily guarantee reciprocity and mutual benefits. Consequently, they could further form small communities that reciprocally exchange information and benefits (P6, P7), where *"more contacts (enabled by these exchanges) bring people closer"* (P7).

As discussed above, beyond acting as virtual bazaar, social commerce further provides novel social interaction opportunities for the traditionally underserved community in Chinese developing regions that suit their particular social and cultural background and practices.

## 5 DISCUSSIONS
## 5.1 Virtual Bazaars and Beyond
As we have shown in our findings, social commerce has demonstrated characteristics resembling but developing beyond offline bazaars, forming an extended version of "virtual bazaar". Specifically, relating to prior literature on bazaars [14, 36, 37], we find that social commerce aligns with bazaars in many aspects: they both support a wide range of items which are mostly small, unbulky, and miscellaneous. Purchases are made point-to-point between sellers and buyers and mostly through small stores or shops without famous brands. Adoptions depend primarily on referrals from strong ties rather than commercial advertisements.

Practices on social commerce also resemble bazaars. Although there is no physical searching efforts on social commerce, peer recommendations are playing similar roles in searching as offline bazaar, helping people identify quality products from available options [14]. Although direct clientilization between customer and seller is less common, a new form of clientilization emerge: trust of the platform resulted from satisfactory purchasing experience lead users to make repeated purchase. Although social commerce uses a publicly marked price system, users have developed their own ways to conduct bazaar-like bargaining. Some social commerce platforms have even devised features with the notion of "bargain", e.g., referring to other users' clicks for lower prices. Although the traditional testing in bazaars does not exist on social commerce because of the physical segregation of sellers and customers, users seek for strong ties' prior testing as references and test product taking advantage of the platform's after-sale services.

Social commerce has also enabled novel collaborative social interactions that fit the community's social, cultural, and economic context. It satisfies the community's collectivism through moving offline "social" shopping online and provides new approaches to earn *mianzi*. The audience choices and behavioral manners of social interactions on social commerce satisfy their needs to maintain *guanxi* and *renqing*. As such, social commerce matches the traditional Chinese socio-cultural norms and elements that are more dominant in the traditionally underserved community in Chinese developing regions, thus further catering to their preferences.

In all aspects, social commerce serves an extended version of bazaars which the traditionally underserved population in Chinese developing regions is accustomed to. This not only makes it possible to transfer their offline experiences online, but also supports a higher level of inclusivity. While a few features such as low prices might be incorporated by other e-commerce models, the unique emphasis on social interactions with users' strong ties, proves to be more successful in engaging the community members in Chinese traditionally underserved community who heavily emphasize the importance of maintaining good relationships with their immediate social circles.

## 5.2 Facilitating Inclusivity
As a novel "counter" case that successfully engages the traditionally underserved community, social commerce manages to suit the traits, appetites, and needs of these people in Chinese developing regions. It thus motivates users to use the platforms, which provides a noteworthy example towards facilitating technology inclusivity. In terms of adoption, social commerce leverages strong ties to penetrate into the traditionally underserved population. Referrals from strong ties let them believe they are qualified for using the platform, and thus more willing to try it out, suiting the community's collectivism culture. Through acting as an extended "virtual bazaar", social commerce further facilitates inclusivity for the community. The resemblance of social commerce to bazaars, which the community is familiar with, not only suits the community's social, economic, and cultural needs, but also effectively lowers the learning cost of the platforms. As such, they can more easily engage with the platforms and avoid non-use, which most mainstream technology experiences. Moreover, as we have shown in Section 4.2, social commerce enables the community to access product not locally available and pursue more cost-efficient options, which further stimulates their platform usages. As a result, the vast traditionally underserved population in Chinese developing regions who is traditionally marginalized and excluded from mainstream technology, actively engages their social economic lives over social commerce and enjoys its convenience and benefits.

### 5.3 Situating into Existing Literature

Our work reflects on, complements, and extends existing literature on HCI4D. Firstly, we echo prior HCI4D literature on the importance of situating into users' real-world practices and value systems [23, 49, 85]. Social commerce manages to accommodate the social, cultural, historical, and economic contexts of Chinese developing regions, thus successfully penetrates into the lives of the traditionally underserved community. This highlights the integration of design practices into locality to guarantee the effectiveness of HCI endeavors [3, 4].

Secondly, we extend prior work by providing a "counter" case for the HCI4D community. Although abundant studies have been conducted on user technology and use experiences in developing regions, they mostly focus on how the platforms fail to consider concerns of the marginalized communities (e.g., [15, 52]). We contribute in a reverse direction by demonstrating how social commerce, as a family of successfully industrially implemented platforms with growing popularity, manages to engage the traditionally underserved community in Chinese developing regions and adapt to their traits and interests. As such, we contribute to the HCI4D community in stressing that we should not only focus on and address obstacles and concerns encountered when marginalized communities' contexts and values are not sufficiently accounted for; but we should also actively learn from the successful experiences of platforms that penetrate into the traditionally underserved and marginalized population.

Thirdly, we complement existing research through a novel case in Chinese developing regions. Although Asian regions have widely attracted the attention of HCI4D scholars [23], HCI4D papers within the Chinese context are rather limited despite the population and size of China [81, 91]. We seek to make up for this research gap by providing an in-depth analysis of social commerce usages of the traditionally underserved community in Chinese developing regions. Specifically, we extend prior HCI4D literature to circumstances of Chinese developing regions where collectivist culture and social interplay are essential in people's decision-making process. We show how the accommodation to the community's accustomed dependencies on strong ties and highlights on social constructs such as *mianzi*, *guanxi*, and *renqing*, facilitates their willingness and comfort of technology usages, which we deem it vital for future HCI design to take into account to actively engage the community through social features. As such, we contribute to the HCI community by 1) responding to the call on broadening the scope of HCI4D to engage with a wider population range [23]; and 2) giving rise to a more thorough understanding of the broader global realities for HCI researchers to consider and adapt to in future design. This could be especially noteworthy considering the possibilities of the occurrence of similar socio-cultural circumstances in eastern culture.

### 5.4 Design Implications

There are several lessons designers and practitioners could draw from social commerce in building more inclusive applications for the traditionally underserved community.

**Take advantage of established habits and infrastructure.** One primary reason that makes it hard for the marginalized community to adopt new technology is the barrier to learn new technology. As suggested by prior literature (e.g., [15]) and our participants, the community is generally not confident about their ability and even sometimes afraid of new technology, and thus tends to avoid trying anything new. One way to get around this fear of new technology among the traditionally underserved community, as demonstrated by social commerce, is to make good use of established habits and infrastructure. For instance, social commerce bases on the infrastructure of WeChat [76] – the most popular instant messaging application in China which has successfully diffused into developing regions already – where users share and post product information in WeChat groups [11]. Given the population is already familiar with WeChat that they have a good knowledge of, it is much easier and natural for them to proceed and try instant messaging based social commerce in comparison to learning other applications anew, thus effectively lowering the "entry bar" for the population.

**Leverage peer influence.** Furthermore, social commerce demonstrates the power of word of mouth and peer influence. As discussed in Section 4.1, peer influence can be central in adopting new technology for the traditionally underserved community: they could not only receive help from them in learning and adopting the new technology, boost their confidence and alleviate "fear of technology", but also result in trust of the technology, which could become a more helpful and cost-efficient way of influence compared to advertising. This method may be especially helpful in communities with strong collectivism culture that would heavily refer to others for decision making.

**Use appropriate metaphors.** Moreover, appropriate metaphors could be a useful strategy to help the traditionally underserved community get to use and engage on application better. As a commonly used design tool, a metaphor is identified as helpful to link technology design to real-world objects, to enhance familiarity, and thus to promote comprehension [27, 41]. In the case of social commerce, the applications use metaphors such as "shop" and "bargain", in the interface design instead of terms that are less familiar to the population as on other platforms. With concrete real-world reflections that they are familiar with, participants could thus understand how to use the platform using their life experience. We thus argue the use of appropriate metaphors could serve as another way to lower technology barriers and promote inclusion through familiar objects and concepts that the population is well-acquainted with.

**Incorporate incentives that suit the community's social, cultural, and economic context.** Finally, practitioners could motivate users with incentives that are suited to their particular social, cultural, and economic context. Under the context of Chinese developing regions, the community is generally quite sensitive about pricing and would be willing to try new things to achieve the goal, which social commerce makes use of through monetary incentives via group buying. The population also looks for ways to maintain relationship with their peers, which social commerce accommodates by enabling novel social interactions. We argue that designers and practitioners should carefully investigate the situated needs of a community and find ways to incentivize the community through creating novel use cases that fit community needs.

### 5.5 Limitations and Future Work

We are aware that there are limitations that our work can suffer from. Firstly, although we ensure that all our interviews are recruited from Chinese cities no higher than tier 4, we are aware that factors such as socio-economic context, literacy rate, Internet penetration, and smartphone usage could vary considering the concrete regional circumstances. Therefore, our study may not cover the specific situations of all developing regions in China. Secondly, our study is conducted in the context of China, where the society and culture can deviate much from the western culture or even other non-western culture (e.g., Indian culture). Therefore, the interplay between certain social, cultural, and economic behaviors could hardly be understood or regarded as appropriate unless considering the Chinese circumstance. Thirdly, our work only focuses on experiences of one stakeholder, *i.e.*, users, on social commerce in developing regions in China. Future work could consider investigating how other stakeholders such as shopkeepers and startups involve with and adapt to the digitized scenarios of social commerce and the complex interplay between offline bazaars and digitized scenarios, as well as how different stakeholders interact with each other. Fourthly, there is the possibility that our findings could suffer from participation bias and selection bias. Users who have more positive experiences on social commerce are possibly more eager to engage in our studies, and we find that our participants also seldom mention the downsides of social commerce. Bearing these aspects in mind, we hold that our discoveries may not apply to other populations and we restrict the direct generalization of any of our findings. However, we insist that the strength of our qualitative work should guarantee a nuanced and in-depth understanding, and our discoveries towards the "counter" case should provide novel implications. Future work could consider large-scale analysis to further verify the mechanisms and findings and their benefits to the traditionally underserved population quantitatively. Future work could also delve into downsides and potential improvements of social commerce to further enlighten design for better experiences.

## 6 CONCLUSION

In this paper, we present an in-depth analysis of the "counter" case on Chinese social commerce, which successfully engages the traditionally underserved community in Chinese developing regions and achieves a great success. We investigate how social commerce enables this high level of inclusivity by qualitatively investigating these people's adoption, product purchases, economic transactions, and social interactions on social commerce. We identify that social commerce acts as and beyond bazaars that the community is familiar with for ages. We illustrate how social commerce adapts to the characteristics of the traditionally underserved community, which facilitates their willingness of engagement and provides satisfying services for them. Based on these, we discuss how our findings can implicate future research and design and contribute to more inclusive HCI systems.

## ACKNOWLEDGMENTS

The authors thank Jinghua Piao for her helpful suggestions on polishing the work. This work was supported in part by the National Natural Science Foundation of China under U1936217, 61971267, 61972223, 61941117, 61861136003.